\documentstyle[aps,preprint,epsfig]{revtex}

\newsavebox{\hflrar}
\sbox{\hflrar}{\makebox[0pt][l]
{${\scriptstyle \leftharpoonup}$}{${\scriptstyle \rightharpoonup}$}}

\def \to {\rightarrow}

\begin{document}

\begin{flushright}
AS-ITP-2003-002
\end{flushright}
\pagestyle{plain}
\vskip 10mm
\begin{center}
{\bf\Large A QCD Analysis of Quark Recombination for Leading Particle Effect } \\
\vskip 10mm
Chao-Hsi Chang, Jian-Ping Ma   \\
{\small {\it Institute of Theoretical Physics , Academia
Sinica, Beijing 100080, China }} \\
~~~ \\
Zong-Guo Si \\
{\small {\it Department of Physics, Shandong University, Jinan Shandong 250100, China}}
\end{center}

\vskip 0.4 cm

\begin{abstract}
The quark recombination mechanism is proposed to explain the
asymmetry between production rates of $D^+$ and of $D^-$ in
their inclusive productions, and also asymmetries for other charmed hadrons.
These asymmetries are observed in experiment and are called as leading particle
effects. In this work
we give a general analysis for contributions of quark recombination
to these asymmetries. The contributions consist of a perturbative- and
nonperturbative part.
We perform two types of factorization by considering the produced hadron
with large transverse momentum  and by taking charm quark as
a heavy quark, respectively. In the case of large transverse momentum
the effect of quark recombination is the standard twist-4 effect. We find
that the contributions are parameterized with four nonperturbative
functions, defined with four quark operators at twist-4, for initial
hadrons without polarization. By taking charm quark as a heavy quark
the factorization can be performed with the heavy quark effective theory(HQET).
The effect of quark recombination is in general parameterized by eight
nonperturbative parameters which
are defined as integrals with matrix elements of four quark operators
defined in HQET. For unpolarized hadrons in the initial state,
the parameters can effectively be reduced to four.
Perturbative parts in the two types of factorization
are calculated at tree-level.
\par\noindent
\end{abstract}

\vfill\eject\pagestyle{plain}\setcounter{page}{1}
\noindent
{\large\bf 1. Introduction}
\par\vskip20pt
Because nonperturbative physics of QCD is still unclear, a prediction
of inclusive hadroproduction of a hadron can not be made in general.
However, if the produced hadron has a large transverse momentum $k_t$,
one can predict the differential cross section. According to the
factorization theorem in QCD(see e.g.,\cite{FaQCD}),
the prediction takes a form as a convolution
with parton distributions for partons in the initial hadrons,
fragmentation function of a produced parton decaying into the produced
hadron, and perturbative functions for partonic processes. The
prediction is based on the leading terms in the so called twist expansion.
In the leading terms twist-2 operators are used to define nonperturbative parts,
i.e., parton distributions and fragmentation functions. It can be shown that
the next-to-leading orders are suppressed by certain power of $k_t$ relatively
to the leading order. At the next-to-leading order, i.e., at twist-4 level,
various partonic processes can happen. The nonperturbative effects are
parameterized by matrix elements of twist-4 operators. At this order,
a produced meson can be formed by recombination of two partons in a initial hadron.
This is the so called parton recombination model\cite{DH},
which is proposed to predict the inclusive production of a meson with low $k_t$.
The recombination can also happen with one of the two
parton involved in a partonic process. This has been studied in \cite{BGS}.
\par
If we take the produced hadron containing one heavy quark in a hadron
collision, for example,$D^+$, and neglect possible $c$-quark content
in the initial hadrons,
then at leading order of perturbative QCD the $D^+$ production can be understood
as that a $c\bar c$ pair is produced via partonic processes like $gg\to c\bar c$
or $q\bar q \to c\bar c$ and then the $c$-quark fragments into the $D^+$ meson.
It is clear that the production rate for $D^+$ and for $D^-$ is the same
at this order. Experimentally significant asymmetries between these production
rates for a class of c-flavored hadrons have been observed and have been called
as leading particle effects. These effects are observed in fixed-target
hadroproduction experiments for the production of c-flavored hadrons
\cite{h1, h2, h3, h4, h5}
and also in photoproduction experiments\cite{p1,p2,p3}. Significant asymmetries
are observed for $D^+-D^-$, $D^0-\bar D^0$ and $D^{*+}-D^{*-}$, while
for $D_s^+$ and $\Lambda_c^+$ the asymmetries with large errors
are consistent with zero. These asymmetries are large when
$k_t$ is small, for $D^+$ and $D^-$ it can be $\sim 0.7$
in the forward direction in hadroproduction.
Theoretically, various explanations exist
\cite{e1,e2}, e.g., these asymmetries can be generated by intrinsic charm
in the initial hadrons, but the predicted asymmetries are smaller
than those in experiment. It should be noted that
these asymmetries can also be generated at next-to-leading
order of perturbative QCD\cite{pq1,pq2,pq3,pq4}, the predicted asymmetries
are too small to explain the observed.
\par
Recently, it has been suggested that these leading particle effects can be
explained with quark recombination\cite{B1,B2,B3}. In E791 experiment\cite{h1}
a $\pi^-$ beam is scattered on a nuclear target, a $D^-$ meson can be produced
through quark recombination of a produced $\bar c$-quark with the valence quark $d$
in $\pi^-$ after hard parton scattering, while a $D^+$ meson can be produced
through quark recombination of a produced $c$-quark with
the sea quark $\bar d$ in $\pi^-$. Because the distribution of $d$- and $\bar d$
quark in $\pi^-$ are significantly different, it results in that more $D^-$ are
produced than $D^+$ in the forward direction. Motivated by NRQCD
factorization\cite{NRQCD}, the authors of \cite{B1,B2,B3} take a factorization
approach for quark recombination, in which a $c$-quark combines with a $\bar d$ quark
in the case for $D^+$ meson and the combination $c\bar d$ is
in $^1S_0$ or $^3S_1$ state, i.e., the pair $c\bar d$ is formed as a
pseudoscalar state or vector state, the $c\bar d$ is then inclusively
transformed into
the $D^+$. If the $\bar d$ quark was a heavy quark, the above
states would be dominant in the quark recombination.
However, the $\bar d$ quark is a light quark, for it
a NRQCD description is not correct. It is possible that the $c\bar d$
is produced in a scalar, pseudovector and tensor state and then transformed
into $D^+$.
\par
If one takes $c$-quark as a heavy quark, a factorization different than that
mentioned at the beginning can be performed
with the Heavy Quark Effective Theory(HQET)\cite{HQET}. In the heavy quark
limit $m_c\to\infty$, the c-flavored hadron produced in a collision will
carry the most momentum of the produced $c$-quark which is transformed
into the hadron after its production. This suggests a factorization
for the production of the c-flavored hadron,  and
in this factorization
the production rate can be written as a product of the production rate of
$c$-quark and a matrix element defined in HQET, corrections to this are suppressed
by certain power of $m_c^{-1}$. This factorization was used to predict heavy quark
fragmentation functions\cite{MaF} and spin alignment of heavy meson in its inclusive
production\cite{MaS}. With this factorization predictions for these two cases
are in good agreement with experiment at $e^+e^-$ colliders\cite{MaF,MaS}.
\par
In this work we will take this factorization to analyze the
possible source for generating leading particle effects. We find
that in general the $c\bar d$ pair can be first formed into the
states of scalar, pseudoscalar, vector and pseudovector, from
these states a $D^+$ can be generated by emitting unobserved
states. Our final results consist of 8 nonperturbative parameters,
which are defined with matrix elements of HQET. Corrections to our
results are suppressed by the inverse of heavy quark mass. These
parameters can be calculated with nonperturbative methods. For
initial hadrons without polarizations, these 8 parameters can be
effectively reduced to 4 because some perturbative coefficients of
the parameters in their contributions to the production rate are
same. If the initial hadrons are polarized, all perturbative
coefficients can be different, the number of parameters can not be
reduced. In the approach of \cite{B1,B2,B3} the contributions from
quark recombination for unpolarized initial state also contain 4
nonperturbative parameters,  but their interpretation is different
than our 4 parameters obtained from the effective reduction of the
8 parameters. Our results hold not only for low $k_t$, but also
for high $k_t$. If $k_t$ is enough large so that all quark masses
can be neglected, the standard factorization mentioned at the
beginning can be performed. In this work we also perform an
analysis for this case. This is similar to the analysis of twist-4
effects in deeply inelastic scattering \cite{twist4}, however, our
task here is simple, because we only need to analyze the
contributions from twist-4 operators, which are defined with
4-quark operators. These twist-4 operators are corresponding to
those 4-quark operators in deeply inelastic scattering. In our
final results for large $k_t$, there are four nonperturbative
functions defined with 4-quark operators separating along the
light-cone. It should be noted that it is rather difficult to
perform a factorization for the relevant processes, in which
hadrons are in the initial- and final state. We will assume that
the factorization holds and especially that the parton model can
be used for initial hadrons. With this assumption we need only to
consider inclusive production of a hadron in a parton collision
and focus on the final hadron.
\par
Our work is organized as the following: In Sect. 2. we give our notations
and analyze the quark recombination in the case with large $k_t$ where the
$k_t$ is so large that all quark masses can be neglected. The contribution
to the differential cross-section is a convolution of perturbative functions
for partonic process with nonperturbative functions defined with four quark
operators. In Sect.3. we analyze the quark recombination with the mentioned
factorization with HQET, in which nonperturbative parts are defined
as matrix elements of HQET. There are eight nonperturbative parameters
characterizing different states of a quark pair which is transformed into
the produced hadron. We also show in detail that
these 8 parameters in the case with unpolarized beams can be effectively
reduced to 4.
Discussions of our results are given. Sect.4 is our
summary.

\par\vskip20pt\noindent
{\bf 2. Quark Recombination at Large $k_t$ and Twist-4 Effect}
\par\vskip20pt
We will denote a heavy quark as $Q$ and a heavy meson containing
one heavy quark $Q$ as $H_Q$. The heavy meson $H_Q$ has a valence quark $\bar q$.
As discussed in the introduction we will consider the production of $H_Q$ in a
parton collision and a initial parton is a valence parton of $H_Q$. Neglecting
possible heavy quark content in initial hadrons, the relevant parton process
for quark recombination is:
\begin{equation}
g(p_1)+ \bar q (p_2) \to H_Q(k) +X,
\end{equation}
where momenta are given in the brackets. We can always divide the unobserved state
into a nonperturbatively produced part $X_N$ and a perturbatively produced
part $X_P$, i.e., $X=X_N+X_P$. At leading order of $\alpha_s$,
$X_P$ is just a antiquark $\bar Q$.
The scattering amplitude for the quark
recombination can be written:
\begin{equation}
{\cal T} = \int \frac{d^4k_1}{(2\pi)^4} A_{ij}(p_1,p_2,k_1,p_3)
              \int d^4x_1 e^{-ik_1\cdot x_1} \langle H_Q +X_N\vert
               \bar Q_i(x_1) q_j(0) \vert 0\rangle,
\end{equation}
where the indices $i,j$ stand for color- and spin indices.
$\bar Q(x_1)$ and $q(0)$ are the fields of $Q$ and $q$ respectively.
$A_{ij}$
is the amplitude for $g(p_1)+ \bar q (p_2) \to Q^*(k_1) + \bar q^*(k_q)
+\bar Q(p_3)$, $^*$ means that the states are not on-shell. If
the states are on-shell, then the amplitude is $\bar u(k_1)_i v_j(k_q) A_{ij}$ with
on-shell conditions for corresponding partons. The Feynman diagrams for the amplitude
$A_{ij}$ are given in Fig.1. For the process we define the variables:
\begin{equation}
 \hat s =(p_1+p_2)^2,\ \ \ \ \ \ \hat t =(p_2-k)^2.
\end{equation}
\par

\begin{center}
\includegraphics[width=10cm]{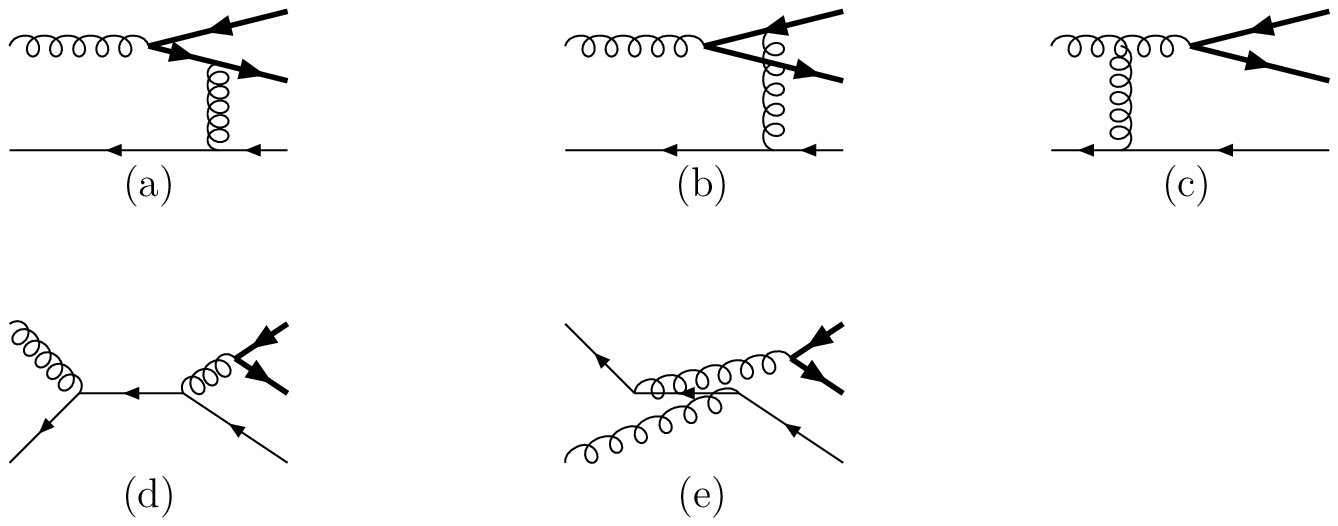}
\centerline{Fig.1. Feynman diagrams for the amplitude $A$. }
\label{Feynman-dg1}
\end{center}
\par\vskip 20pt
The contribution from the process in Eq.(1) to the differential cross section is:
\begin{eqnarray}
 d\sigma_R &=&\frac{1}{2\hat s} \sum_{X_N}
    \frac {d^3k}{(2\pi)^32k^0}\int \frac{d^3p_3}{(2\pi)^32p_3^0}
       (2\pi)^4\delta^4(p_1+p_2-k-p_3-P_{X_N})\nonumber\\
      &&\cdot  \int \frac{d^4k_1}{(2\pi)^4} \frac{d^4k_2}{(2\pi)^4}
         A_{ij}(p_1,p_2,k_1,p_3) (\gamma^0 A^\dagger
         (p_1,p_2,k_2,p_3)\gamma^0)_{kl} \nonumber\\
      &&\cdot \int d^4x_1 d^4x_2 e^{-ik_1\cdot x_1+ik_2\cdot x_2}
      \langle 0\vert \bar q_k(0) Q_l(x_2) \vert H_Q+X_N\rangle
      \langle H_Q +X_N\vert \bar Q_i(x_1) q_j(0) \vert 0\rangle,
\end{eqnarray}
where we use the subscribe $R$ to denote the contribution from quark recombination,
the average of spin and color of initial partons is implied.
If $H_Q$ has nonzero spin, the summation over the spin is understood.
Using translational covariance one can eliminate the sum over $X_N$.
We define $a_{H_Q}^\dagger$ as the creation operator for $H_Q$ and we
obtain:
\begin{eqnarray}
 d\sigma_R &=& \frac{1}{2\hat s}\frac {d^3k}{(2\pi)^32k^0}
         \int \frac{d^3p_3}{(2\pi)^32p_3^0}
      \cdot  \int \frac{d^4k_1}{(2\pi)^4} \frac{d^4k_2}{(2\pi)^4}
         A_{ij}(p_1,p_2,k_1,p_3) (\gamma^0 A^\dagger
         (p_1,p_2,k_2,p_3)\gamma^0)_{kl} \nonumber\\
      &&\cdot \int d^4x_1 d^4x_2 d^4x_3 e^{-ik_1\cdot x_1+ik_2\cdot x_2
       -ix_3\cdot k_3}
      \langle 0\vert \bar q_k(0) Q_l(x_2)  a^\dagger _{H_Q}(k)
       a_{H_Q}(k) \bar Q_i(x_1) q_j(x_3) \vert 0\rangle,
\end{eqnarray}
with $k_3=p_1+p_2-k_1-p_3$ as the momentum of $\bar q$ after the hard scattering
in the amplitude $A_{ij}(p_1,p_2,k_1,p_3)$.
This contribution can be conveniently represented by the diagram in Fig.2.,
where the lower part represents the perturbative part, i.e.,
$A_{ij}(p_1,p_2,k_1,p_3) (\gamma^0 A^\dagger
(p_1,p_2,k_2,p_3)\gamma^0)_{kl}$, the upper part with the black box represents
the nonperturbative part, i.e., the Fourier transformed matrix element in the second
line of the above equation.
\par\vskip20pt

\begin{center}
\includegraphics[width=8cm]{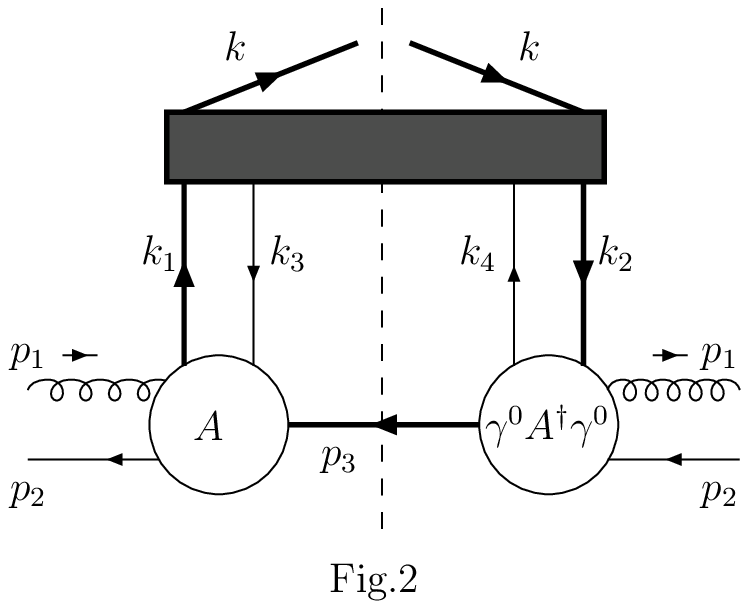}
\end{center}
{Fig.2. Diagram representation for Eq.(5). In the figure
$k_4=k_1-k_2+k_3$, the dash line is the cut, the thick lines below the black
box are for heavy quarks. }
\label{Feynman-dg1}

\par\vskip 20pt
\par
The factorization can be performed by decomposing the color- and Dirac indices of
the matrix element.
The decomposition of color indices of the matrix element can be performed
in a straightforward way and one can identify that the $Q\bar q$ pair is
in a color-singlet or a color-octet. We denote the color-singlet- and color-octet
contribution as $d\sigma_R^{(1)}$ and $d\sigma_R^{(8)}$ respectively and
\begin{equation}
d\sigma_R =d\sigma_R^{(1)} + d\sigma_R^{(8)}.
\end{equation}
The contributions read:
\begin{eqnarray}
 d\sigma_R^{(1)} &=& \frac{1}{2\hat s}
      \frac {d^3k}{(2\pi)^32k^0}\int \frac{d^3p_3}{(2\pi)^32p_3^0}
      \cdot  \int \frac{d^4k_1}{(2\pi)^4} \frac{d^4k_2}{(2\pi)^4}
       \nonumber\\
      && \cdot \frac{1}{9} {\rm Tr} A_{ij}(p_1,p_2,k_1,p_3) {\rm Tr}(\gamma^0 A^\dagger
         (p_1,p_2,k_2,p_3)\gamma^0)_{kl}
       \int d^4x_1 d^4x_2 d^4x_3
       \nonumber\\
       && \cdot e^{-ik_1\cdot x_1+ik_2\cdot x_2
       -ix_3\cdot k_3}
      \langle 0\vert \bar q_k(0) Q_l(x_2)  a^\dagger _{H_Q}(k)
       a_{H_Q}(k) \bar Q_i(x_1) q_j(x_3) \vert 0\rangle,
\end{eqnarray}
\begin{eqnarray}
 d\sigma_R^{(8)} &=& \frac{1}{2\hat s}
      \frac {d^3k}{(2\pi)^32k^0} \int \frac{d^3p_3}{(2\pi)^32p_3^0}
      \cdot  \int \frac{d^4k_1}{(2\pi)^4} \frac{d^4k_2}{(2\pi)^4}
      \nonumber\\
      &&\cdot \frac{1}{2} {\rm Tr}T^b A_{ij}(p_1,p_2,k_1,p_3)
       {\rm Tr}(T^b\gamma^0 A^\dagger
         (p_1,p_2,k_2,p_3)\gamma^0)_{kl}
       \int d^4x_1 d^4x_2 d^4x_3
      \nonumber\\
       &&\cdot e^{-ik_1\cdot x_1+ik_2\cdot x_2
       -ix_3\cdot k_3}
      \langle 0\vert \bar q_k(0)T^a Q_l(x_2)  a^\dagger _{H_Q}(k)
       a_{H_Q}(k) \bar Q_i(x_1)T^a q_j(x_3) \vert 0\rangle.
\end{eqnarray}
In the above equations the traces are taken over color indices and
the indices $i,j,k,l$ stand only for Dirac indices.
\par
In this section we will consider the case where the transverse momentum $k_t$ of
the produced hadron is very large, and
all quark masses can be neglected.
To evaluate the contributions we take a frame in which
the $z$-direction is the moving direction of $H_Q$ and define
a light-cone coordinate system in which a vector $V$ is expressed with components
$V^\mu (\mu =0,1,2,3)$
in usual coordinate system as:
\begin{equation}
V^\mu =(V^+,V^-,V^1,V^2)=\left ( \frac{V^0+V^3}{\sqrt{2}},
   \frac{V^0-V^3}{\sqrt{2}},V^1,V^2 \right ).
\end{equation}
We also introduce two light-cone vectors:
\begin{equation}
 n^\mu =(0,1,0,0),\ \ \ \ \, l^\mu =(1,0,0,0).
\end{equation}
The momentum $k$ reads:
\begin{equation}
k^\mu = (k^+,k^-,0,0) = (k^+, \frac{M_{H_Q}^2}{2k^+},0,0)\approx  (k^+,0,0,0)=k^+l^\mu.
\end{equation}
\par
The $x$-dependence of the matrix elements in Eq.(7) and Eq.(8) is controlled by different
scales. The $x^-$ dependence is controlled by the large scale $k^+$, while
the $x^+$-dependence is controlled by the small scale $k^-$ or $\Lambda_{QCD}$,
the $x^1-$ and $x^2$-dependence is controlled by $\Lambda_{QCD}$. Therefore, as
an approximation one can neglect the dependence of $x^-$, $x^{1}$ and $x^{2}$.
Corrections to this can be systematically added. This results in the so called
collinear expansion. The leading order of the expansion is to set
\begin{equation}
k_i^\mu =(z_ik^+,0,0,0)=z_ik^+ l^\mu, \ \ \ {\rm for}\ i=1,2,3,
\end{equation}
in the amplitude $A$ and $A^\dagger$.
Taking the color-singlet in Eq.(7) as an example, we obtain:
\begin{eqnarray}
 d\sigma_R^{(1)} &=& \frac{1}{2\hat s} \frac {d^3k}{(2\pi)^32k^0} \int\frac {d^3p_3}{(2\pi)^32p_3^0}
      \cdot  \int \frac{dz_1}{(2\pi )} \frac{dz_2}{(2\pi )} (k^+)^2
       (2\pi)^3\delta_-^3(p_1+p_2-k_1-p_3) \nonumber\\
       && \frac{1}{9} {\rm Tr} A_{ij}(p_1,p_2,z_1k^+ l,p_3)
        {\rm Tr}(\gamma^0 A^\dagger
         (p_1,p_2,z_2 k^+l,p_3)\gamma^0)_{kl} \nonumber\\
      &&\cdot \int dx_1^- dx_2^- dx_3^- e^{-ik^+(z_1 x_1^- -z_2 x_2^-
       +z_3x_3^-)}\nonumber\\
      && \langle 0\vert \bar q_k(0) Q_l(x_2^- n)  a^\dagger _{H_Q}(k)
       a_{H_Q}(k) \bar Q_i(x_1^- n) q_j(x_3^- n) \vert 0\rangle,
\end{eqnarray}
where $\delta_-^3(q) = \delta(q^-)\delta(q^1)\delta(q^2)$.
With Eq.(12) the variable $z^{-1}_i(i=1,2,3)$ is just the momentum fraction
of partons carried by $H_Q$. Because the momentum conservation they are bounded,
for example, $z_i >1$ for $i=1,2,3$. Now the decomposition of the Dirac indices
can be performed. We will only keep the leading terms, i.e., the twist-4 terms.
The decomposition read:
\begin{eqnarray}
 && k^+\int \frac{dx_1^-}{(2\pi)}\frac{dx_2^-}{(2\pi)}\frac{dx_3^-}{(2\pi)}
       e^{-ik^+(z_1 x_1^- -z_2 x_2^-
       +z_3x_3^-)}
\langle 0\vert \bar q_k(0) Q_l(x_2^- n)  a^\dagger _{H_Q}(k)
       a_{H_Q}(k) \bar Q_i(x_1^- n) q_j(x_3^- n) \vert 0\rangle \nonumber\\
 &=& (\gamma^-)_{ji}(\gamma^-)_{lk}{\cal T}_1^{(1)}(z_1,z_2,z_3)
      +(\gamma^-\gamma_5)_{ji}(\gamma^-\gamma_5)_{lk}
      {\cal T}_2^{(1)}(z_1,z_2,z_3)
      \nonumber\\
    && +(i\sigma^{-m})_{ji}(i\sigma^{-m})_{lk}
      {\cal T}_3^{(1)}(z_1,z_2,z_3) +\cdots,
\end{eqnarray}
where $\cdots$ denotes the terms with twist higher than 4. The sum
over the index $m$ runs from 1 to 2. The functions ${\cal T}_i(i=1,2,3)$
are defined as:
\begin{eqnarray}
{\cal T}_1^{(1)}(z_1,z_2,z_3) &=&
  \frac{k^+}{16}\int \frac{dx_1^-}{2\pi}\frac{dx_2^-}{2\pi}\frac{dx_3^-}{2\pi}
       e^{-ik^+(z_1 x_1^- -z_2 x_2^-
       +z_3x_3^-)}\nonumber\\
 &\cdot &\langle 0\vert \bar q(0) \gamma^+ Q(x_2^- n)  a^\dagger _{H_Q}
       a_{H_Q} \bar Q(x_1^- n) \gamma^+ q(x_3^- n) \vert 0\rangle,
       \nonumber\\
{\cal T}_2^{(1)}(z_1,z_2,z_3) &=&
  \frac{k^+}{16}\int \frac{dx_1^-}{2\pi}\frac{dx_2^-}{2\pi}\frac{dx_3^-}{2\pi}
       e^{-ik^+(z_1 x_1^- -z_2 x_2^-
       +z_3x_3^-)}\nonumber\\
  &\cdot& \langle 0\vert \bar q(0) \gamma^+\gamma_5 Q(x_2^- n)  a^\dagger _{H_Q}
       a_{H_Q} \bar Q(x_1^- n) \gamma^+\gamma_5 q(x_3^- n) \vert 0\rangle,
  \nonumber\\
  {\cal T}_3^{(1)}(z_1,z_2,z_3) &=&
  \frac{k^+}{16}\int \frac{dx_1^-}{2\pi}\frac{dx_2^-}{2\pi}\frac{dx_3^-}{2\pi}
       e^{-ik^+(z_1 x_1^- -z_2 x_2^-
       +z_3x_3^-)} \nonumber\\
 &\cdot& \langle 0\vert \bar q(0) i\sigma^{+i}Q(x_2^- n)  a^\dagger _{H_Q}
       a_{H_Q} \bar Q(x_1^- n) i\sigma^{+i} q(x_3^- n) \vert 0\rangle,
\end{eqnarray}
these three functions represent the  nonperturbative effects related
to the produced hadron. They are invariant under a Lorentz boost along
the moving direction of $H_Q$. The dimension of these functions
are 2 in mass, hence they are proportional to $\Lambda^2_{QCD}$.
With these results we obtain
\begin{eqnarray}
 d\sigma_R^{(1)} &=& \frac{1}{2\hat s}
     \frac {d^3k}{(2\pi)^22k^0} \int\frac {d^3p_3}{(2\pi)^32p^0_3}
      \cdot  \int \frac{dz_1}{z_1} \frac{dz_2}{z_2}
       (2\pi)^3\delta_-^3(p_1+p_2-k_1-p_3) \nonumber\\
       && \cdot\frac{1}{9k^+} \big\{ {\rm Tr} (\gamma\cdot k_1 A(p_1,p_2,k_1,p_3))
        {\rm Tr}(\gamma\cdot k_2\gamma^0 A^\dagger
         (p_1,p_2,k_2,p_3)\gamma^0) \cdot {\cal T}_1^{(1)}(z_1,z_2,z_3)
          \nonumber\\
       && +  {\rm Tr} (\gamma\cdot k_1\gamma_5 A (p_1,p_2,k_1,p_3))
        {\rm Tr}(\gamma\cdot k_2\gamma_5\gamma^0 A^\dagger
         (p_1,p_2,k_2,p_3)\gamma^0) \cdot {\cal T}_2^{(1)}(z_1,z_2,z_3)
         \nonumber\\
      && + {\rm Tr} (i\sigma^{\mu i} k_{1\mu}  A(p_1,p_2,k_1,p_3))
        {\rm Tr}(i\sigma^{\mu i} k_{2\mu} \gamma^0 A^\dagger
         (p_1,p_2,k_2,p_3)\gamma^0) \cdot {\cal T}_3^{(1)}(z_1,z_2,z_3)
         \big\},
\end{eqnarray}
for the color-octet contributions we obtain:
\begin{eqnarray}
 d\sigma_R^{(8)} &=& \frac{1}{2\hat s}\frac {d^3k}{(2\pi)^22k^0} \int\frac {d^3p_3}{(2\pi)^32p_3^0}
      \cdot  \int \frac{dz_1}{z_1} \frac{dz_2}{z_2}
       (2\pi)^3\delta_-^3(p_1+p_2-k_1-p_3) \nonumber\\
       &&\cdot \frac{1}{2k^+} \big\{ {\rm Tr} (\gamma\cdot k_1T^a A(p_1,p_2,k_1,p_3))
        {\rm Tr}(\gamma\cdot k_2\gamma^0 T^a A^\dagger
         (p_1,p_2,k_2,p_3)\gamma^0) \cdot {\cal T}_1^{(8)}(z_1,z_2,z_3)
          \nonumber\\
       && +  {\rm Tr} (\gamma\cdot k_1\gamma_5 T^a A (p_1,p_2,k_1,p_3))
        {\rm Tr}(\gamma\cdot k_2\gamma_5\gamma^0 T^a A^\dagger
         (p_1,p_2,k_2,p_3)\gamma^0) \cdot {\cal T}_2^{(8)}(z_1,z_2,z_3)
         \nonumber\\
      && + {\rm Tr} (i\sigma^{\mu i} k_{1\mu}  T^a A(p_1,p_2,k_1,p_3))
        {\rm Tr}(i\sigma^{\mu i} k_{2\mu} \gamma^0 T^a A^\dagger
         (p_1,p_2,k_2,p_3)\gamma^0) \cdot {\cal T}_3^{(8)}(z_1,z_2,z_3)
         \big\},
\end{eqnarray}
with $k_i=z_i k^+ l$.
The definitions for the functions ${\cal T}_i^{(8)}(z_1,z_2,z_3)(i=1,2,3)$
can be obtained by replacing the color matrix $1\otimes 1$ with
$T^a \otimes T^a$ in the definitions for
${\cal T}_i^{(1)}(z_1,z_2,z_3)(i=1,2,3)$, respectively.
\par
The integration over phase space can be easily performed in the
center-of-mass frame. Consider the integral
\begin{eqnarray}
  \int\frac {d^3k}{(2\pi)^32k^0}
 \int\frac {d^3p_3}{(2\pi)^32p_3^0}
      \cdot
       (2\pi)^3 \delta_-^3(p_1+p_2 -k_1-p_3)\cdot\frac{1}{k^+},
\end{eqnarray}
because the only nonzero component of $k_1$ is $k_1^+$, $k_1$ can be set zero
in the $\delta_-$ function. The integration over $p_3$ can be done with the
$\delta_-$ function. We insert in the above integral the identity:
\begin{equation}
\int d z_3 k^+\delta(p_1^++p_+-z_1k^+ -z_3k^+)=1
=\int d z_3\sqrt{2}\vert {\bf k}\vert \delta (
 \frac{\sqrt{\hat s}}{\sqrt{2}} -\sqrt{2}(z_1+z_3)\vert {\bf k}\vert ),
\end{equation}
and obtain the phase space integral as
\begin{equation}
\int\frac {d^3k}{(2\pi)^32k^0}
 \int\frac {d^3p_3}{(2\pi)^32p_3^0}
      \cdot
       (2\pi)^3 \delta_-^3(p_1+p_2 -k_1-p_3)\cdot\frac{1}{k^+}
       = \frac{1}{4(2\pi)^2 \hat s  }
   \int d \hat t \int d z_3 \cdot \frac{1}{(z_1+z_3)^2}.
\end{equation}
With this we can express our results for $d\sigma/d\hat t$. The remainder of the
calculations are straightforward. Without detailed calculation one can find
that the coefficient of ${\cal T}_3^{(i)}$ is zero, the reason is that
we consider unpolarized hadron beams here and the polarizations of
initial partons are averaged. Due to helicity conservation, the coefficient
must be zero. If we consider polarized hadron beams, ${\cal T}_3^{(i)}$ can
lead to a nonzero contribution.
With the structure of ${\cal T}_1^{(i)}$ and ${\cal T}_2^{(i)}$ one can also find
that
the coefficients of ${\cal T}_1^{(i)}$ and
${\cal T}_2^{(i)}$ are same. Finally we have:


\begin{eqnarray}
 \frac{d\sigma_R^{(1)}}{d\hat t}(\hat s, \hat t)
 &=&  \frac{2\pi\alpha_s^31}{243\hat s^2  }
\int \frac{dz_1}{z_1} \,
\frac{dz_2}{z_2}\,\frac{dz_3}{z_3} \,\frac{1}{(z_1+z_3)^2}\,
\frac{ \big\{ {\cal T}_1^{(1)}+{\cal T}_2^{(1)}\big\}}
{(z_1-z_2+z_3)(\hat{s}+\hat{t})} \nonumber\\
&&\Big\{\frac{\hat{s}^2}{\hat{t}^2} (9 z_1+z_3)(z_1+8 z_2+z_3)
+\frac{\hat{s}}{\hat{t}}\Big[27 z_1^2+297 z_1 z_2+29 z_1 z_3+25 z_2 z_3+2 z_3^2\Big]
\nonumber\\
&&+\frac{4\hat{t}^2}{\hat{s}^2}\Big[81 z_3(z_3-z_2)+z_1 (25 z_2+81 z_3)\Big]
-\frac{4\hat{t}}{\hat{s}}\Big[81z_1^2+9z_3(8z_2+z_3)+z_1(90z_3-86z_2)\Big]
\nonumber\\
&&-144 z_1^2+469 z_1 z_2-179 z_1 z_3-109 z_2 z_3-35 z_3^2
\Big\},
\end{eqnarray}

\begin{eqnarray}
 \frac{d\sigma_R^{(8)}}{d\hat t}(\hat s, \hat t) &=&
  \frac{\pi\alpha_s^3}{162\hat s^2 }
\int \frac{dz_1}{z_1} \,\frac{dz_2}{z_2}\,\frac{dz_3}{z_3} \,
\frac{1}{(z_1+z_3)^2} \,
\frac{\big\{ {\cal T}_1^{(8)}+{\cal T}_2^{(8)}\big\}}
{(z_1-z_2+z_3)\Big[\hat{s}+\hat{t}\Big]} \nonumber\\
&&\Big\{\frac{\hat{s}^2}{\hat{t}^2} \Big[63z_1^2+z_3(z_2+62 z_3)+z_1(9z_2+125 z_3)\Big]
\nonumber\\
 &&+\frac{\hat{s}}{\hat{t}}\Big[207z_1^2+99z_1 z_2+331 z_1 z_3+83 z_2 z_3+124 z_3^2\Big]
 \nonumber\\
&&+\frac{7\hat{t}^2}{\hat{s}^2}\Big[45z_3(z_3-z_2)+z_1(41z_2+45 z_3)\Big]
-\frac{4\hat{t}}{\hat{s}}\Big[81z_1^2+9z_3(z_2+8z_3)+z_1(153z_3-148z_2)\Big]
\nonumber\\
&&-18z_1^2+5z_1(79z_2-47z_3)+z_3(199z_2-217z_3)\Big\}.
\end{eqnarray}
Similarly for the process
\begin{equation}
\gamma(p_1)+\bar{q}(p_2)\rightarrow H_Q(k)+X,
\end{equation}
we have
\begin{eqnarray}
 \frac{d\tilde{\sigma}_R^{(i)}}{d \hat t}(\hat s,\hat t) &=&
\frac{\pi\alpha_s^2\alpha}{3\hat s^2 }
\int d \hat{t}
\int \frac{dz_1}{z_1} \,\frac{dz_2}{z_2}\,\frac{dz_3}{z_3} \,
\frac{1}{(z_1+z_3)^2}
\frac{{\cal C}^{(i)} e_Q^2
\big\{ {\cal T}_1^{(i)}+{\cal T}_2^{(i)}\big\}}
{(z_1-z_2+z_3)} \\ \nonumber
&&\Big\{\frac{\Big[\hat{t}^2 z_1 z_2 + \hat{s}^2 z_3 (z_1-z_2+z_3)\Big]}{\hat{t}^2
\left(\hat{s} + \hat{t} \right)}\Big[1+\kappa\frac{\hat{t}}{\hat{s}}\Big]^2\Big\},
\end{eqnarray}
where $e_Q$($e_q$) is the charge of the heavy(light) quark and
\begin{equation}
\kappa=\frac{e_q}{e_Q}, ~~~~~~~
{\cal C}^{(1)}=\frac{256}{27}, ~~~~~~~~ {\cal C}^{(8)}=\frac{8}{9}.
\end{equation}
\par
With these results one can predict the leading particle effect in hadron collision
and photoproduction. We denote the anti-particle of $H_Q$ as $\bar H_Q$. Because
of the symmetry of charge conjugation, the differential cross section for
$g(p_1)+q(p_2)\to \bar H_Q(k) +X$ is the same as given above. The leading particle
effect is generated by the asymmetry between the distributions of $q$ and $\bar q$.
We consider the production in the hadron collision:
\begin{eqnarray}
  A(P_1) +B(P_2) &\to& H_Q(k) +X, \nonumber \\
  A(P_1) +B(P_2) &\to& \bar H_Q(k) +X,
\end{eqnarray}
where $A$ and $B$ are the initial hadrons, whose spin is not observed. We define
the variables:
\begin{equation}
 s= (P_1+P_2)^2, \ \ \  t=(P_2-k)^2,\ \ \ \  u=(P_1-k)^2.
\end{equation}
The leading particle effect can be predicted as:
\begin{eqnarray}
\frac{d\sigma}{dt}(H_Q) - \frac{d\sigma}{dt}(\bar H_Q)
   &=& \int d x_1 d x_2 f_{g/A}(x_1) \left [f_{\bar q/B}(x_2) -
         f_{q/B} (x_2)\right ]\cdot
         \frac{d\sigma_R}{d\hat t}(\hat s=x_1x_2s, \hat t=x_2t)
  \nonumber \\
  &+& \int d x_1 d x_2 f_{g/B}(x_2) \left [ f_{\bar q/A}(x_1) -
         f_{q/A} (x_1) \right ] \cdot
  \frac{d\sigma_R}{d\hat t}(\hat s=x_1x_2s, \hat t=x_1 u),
\end{eqnarray}
where $d\sigma_R$ is the sum of contributions given in Eq.(21) and Eq.(22),
$f_{a/A}(x)$ is the distribution for parton $a$ in hadron $A$. Similarly,
one can also obtain the leading particle effect in photoproduction, replacing
$A$ with a photon in Eq.(26) we obtain:
\begin{equation}
\frac{d\sigma}{dt}(H_Q) - \frac{d\sigma}{dt}(\bar H_Q)
   = \int d x_1 \left [f_{\bar q/B}(x_1) -
         f_{q/B} (x_1)\right ]\cdot
         \frac{d\tilde\sigma_R}{d\hat t}(\hat s=x_1s, \hat t=x_1t),
\end{equation}
where $d\tilde \sigma_R = d\tilde\sigma_R^{(1)}+d\tilde\sigma_R^{(8)}$.
If $A$ is the anti-particle of $B$, then the leading particle effect vanishes
because of the charge conjugation symmetry. Since the nonperturbative
functions ${\cal T}^{(1)}_i$ and ${\cal T}^{(8)}_i$ for $i=1,2$ are unknown,
numerical predictions can not be made. These functions may be studied with
nonperturbative methods or extracted from experimental results, they are universal
like parton distributions. Once they are known, one can make numerical predictions.
Our results here hold for the case with large transverse momentum of the produced
hadron, i.e. For $s\to\infty$ and $t\to\infty$, but with the ratio $t/s$
being fixed, one obtains the power behavior
\begin{equation}
\frac{d\sigma}{dt}(H_Q) - \frac{d\sigma}{dt}(\bar H_Q) \sim \frac{1}{s^3},
\end{equation}
in comparison with the production at twist-2 level
\begin{equation}
\frac{d\sigma}{dt}(H_Q) \sim \frac{d\sigma}{dt}(\bar H_Q) \sim \frac{1}{s^2},
\end{equation}
hence, the leading particle effect through quark recombination is suppressed
in comparison of the production rate
by a factor $s^{-1}$, which is typical for twist-4 effects.
\par
In \cite{BGS} contributions from quark recombination to inclusive $\pi$ production
is studied,
partonic processes like $g\bar q \to \pi g$ and $g q\to q\pi$
are calculated by taking a wave function at leading twist for $\pi$.
This is different than our approach. If one only keeps
the contribution from ${\cal T}^{(1)}_2$ in Eq.(21) and neglects
all other contributions, and then for ${\cal T}^{(1)}_2$ one uses
the approximation of vacuum saturation to write it in term of the $\pi$ wave function
at leading twist, our approach is equivalent
to that in \cite{BGS}.  It is clear that there are no reasons
to neglect other contributions and to use the approximation of vacuum saturation.
In \cite{BGS} the results are given by taking asymptotic form of the wave function,
where the integration over the momentum fraction as the variable
of the wave function is performed analytically. This prevents us from a direct comparison
with our results. From our results it is clear that the effect of quark recombination
is characterized by 2 nonperturbative functions, effectively for unpolarized
beams, while for polarized beams there are in general 6 nonperturbative functions.
\par


\par\vskip20pt
\noindent
{\bf 3. HQET Factorization for Quark Recombination}
\par\vskip20pt
In this section we will use nonrelativistic normalization for $H_Q$ and
heavy quarks $Q$. If the pole mass $m_Q$ of the heavy quark $Q$ is heavy enough,
one can expand
the heavy quark field $Q(x)$
with the field in HQET\cite{HQET}:
\begin{eqnarray}
  Q(x) &=& e^{-im_Qv\cdot x} \cdot\left\{
    h(x) +{\cal O}(m_Q^{-1})\right \}+\cdots, \nonumber \\
  \bar Q(x) &=& e^{+im_Qv\cdot x} \cdot \left\{
    \bar h(x) +{\cal O}(m_Q^{-1})\right \}+\cdots,
\end{eqnarray}
where $v$ is the four velocity of $H_Q$, i.e., $v=k/M_{H_Q}$. $h(x)$ is the
field of HQET and depends on $v$ implicitly,
the $\cdots$ denotes the part of anti-quark. The fields have
the following property:
\begin{equation}
  v\cdot\gamma h(x)=h(x),\ \ \ \, \bar h(x) v\cdot\gamma =\bar h(x).
\end{equation}
For $m_Q\to\infty$
the most momentum of a produced $Q$ is carried by $H_Q$, where the produced
heavy quark $Q$ combines with other light quarks and gluons to form the hadron
$H_Q$. Using this fact and HQET one can expand
the cross-section for production of $H_Q$ in $\Lambda_{QCD}/m_Q$.
Following the proposed factorization in \cite{MaF,MaS}, the production rate
at leading order of $\Lambda_{QCD}/m_Q$ is a product of the production rate of $Q$
with a matrix element defined in HQET, the momentum of $H_Q$ is approximated
by the momentum of $Q$ and $M_{H_Q}\approx m_Q$. For example, the production
of $H_Q$ via gluon fusion, i.e., $g+g\to H_Q +X$, the differential cross-section
can be written as:
\begin{equation}
d\sigma (g+g\to H_Q +X) = d\sigma (g+g\to Q +X)
    \cdot\frac{1}{6}
    \langle 0\vert {\rm Tr} h a^\dagger_{H_Q} a_{H_Q} \bar h \vert 0\rangle.
    \cdot \left\{ 1+{\cal O}(\frac{\Lambda^2_{QCD}}{m_Q^2})\right\},
\end{equation}
where matrix element is defined in the rest frame of $H_Q$. The above
equation has a close correspondence to inclusive decay of $H_Q$ with
the factorization, the corresponding
matrix element for inclusive decay of $H_Q$ equals 1 because flavor conservation
of HQET, it leads to the well known result\cite{HQET}:
\begin{equation}
\Gamma (H_Q\to X) = \Gamma ( Q\to X)
\cdot \left\{ 1+{\cal O}(\frac{\Lambda^2_{QCD}}{m_Q^2})\right\}.
\end{equation}
At the next-to-leading order for the decay width the nonperturbative
effect is represented by dimension 5 operators, corresponding operators
for production can also be found. These operators are bilinear in quark fields.
Going beyond this order, i.e., at the order of $m_Q^{-3}$, one will encounter
four quark operators, these 4-quark operators represent the nonperturbative
effect for quark recombination.
\par
For the contributions from quark recombination,
we neglect higher orders in Eq.(32). The color-singlet contribution
reads:
\begin{eqnarray}
 d\sigma_R^{(1)} &=&\frac{1}{2\hat s}  \frac {d^3k}{(2\pi)^3}
             \int \frac {d^3p_3}{(2\pi)^3}
      \cdot  \int \frac{d^4k_1}{(2\pi)^4} \frac{d^4k_2}{(2\pi)^4}
       \frac{1}{9} {\rm Tr} A_{ij}(p_1,p_2,k_1,p_3) {\rm Tr}(\gamma^0 A^\dagger
         (p_1,p_2,k_2,p_3)\gamma^0)_{kl} \nonumber\\
      &&\cdot \int d^4x_1 d^4x_2 d^4x_3 e^{-ik_1\cdot x_1+ik_2\cdot x_2
       -ix_3\cdot k_3-im_Q v\cdot x_2 +im_Q v\cdot x_1}
       \nonumber\\
     && \cdot \langle 0\vert \bar q_k(0) h_l(x_2)  a^\dagger _{H_Q}(k)
       a_{H_Q}(k) \bar h_i(x_1) q_j(x_3) \vert 0\rangle.
\end{eqnarray}
Now the space-time dependence of the matrix element is controlled by the
small scale $\Lambda_{QCD}$, reflecting the fact that the most momentum
of $Q$ is carried by the hadron $H_Q$, the difference between the momentum
of $Q$ and that of $H_Q$ is order of $\Lambda_{QCD}$. At first look, one may
neglect the space-time dependence as an approximation, used in deriving
Eq.(34) or Eq.(35). This implies that the light antiquark $\bar q$ will
have zero momentum after scattering. It results in that the amplitude $A$
will be divergent because the gluon propagator attached to $\bar q$ in Fig.1.
To have meaningful predictions one can not neglect the space-time dependence
here. The momentum of $\bar q$ is of order of $\Lambda_{QCD}$, one can expand
the amplitude in this momentum, this approach is used in \cite{B1,B2,B3}.
We will use the approach here to regularize the divergence. For doing this
we write the variables $x_i$ as:
\begin{equation}
 x_i^\mu = v\cdot x_i v^\mu +x_{Ti}^\mu
         = \omega_i v^\mu +x_{Ti}^\mu, \ \ \
          {\rm for} \ i=1,2,3,
\end{equation}
momenta can also be decomposed similarly as above.
The dependence of the matrix element on $x_{Ti}(i=1,2,3)$ can be safely neglected,
and we have:
\begin{eqnarray}
 d\sigma_R^{(1)} &=& \frac{1}{2\hat s}
          \frac {d^3k}{(2\pi)^3}\int\frac{d^3p_3}{(2\pi)^3}
        (2\pi)^3\delta^3 (p_{T1}+p_{T2}-p_{T3})
        \nonumber \\
   &&   \cdot  \int \frac{d\eta_1}{2\pi} \frac{d\eta_2}{2\pi}
       \frac{1}{9} {\rm Tr} A_{ij}(p_1,p_2,k_1,p_3) {\rm Tr}(\gamma^0 A^\dagger
         (p_1,p_2,k_2,p_3)\gamma^0)_{kl} \nonumber\\
      &&\cdot \int d\omega_1 d\omega_2 d\omega_3
         e^{-i\eta_1\cdot \omega_1+i\eta_2\cdot \omega_2
       -i\eta_3\cdot\omega_3}
        \langle 0\vert \bar q_k(0) h_l(\omega_2 v)  a^\dagger _{H_Q}(k)
       a_{H_Q}(k) \bar h_i(\omega_1 v) q_j(\omega_3v) \vert 0\rangle,
\end{eqnarray}
with
\begin{equation}
k_1 = (m_Q + \eta_1)v,\ \ \ k_2 =(m_Q +\eta_2) v,\ \ \ \  k_3= \eta_3v.
\end{equation}
To separate the divergences more clearly,
we use translational invariance for the matrix element to shift the variable
of fields with $-\omega_2 v$, and rearrange the variables as:
$\omega_2 \to -\omega_2$, $\omega_3-\omega_2\to \omega_1$,
$\omega_1-\omega_2\to \omega_3$, and $\eta_1+\eta_3-\eta_2\to \eta_2$,
$\eta_1 \to \eta_3$, $\eta_3 \to \eta_1$. After this re-arrangement
the momentum $k_i$ reads:
\begin{equation}
k_1 = (m_Q + \eta_3)v,\ \ \ k_2 =(m_Q +(\eta_3+\eta_1-\eta_2)) v,\ \ \ \
k_3= \eta_1v.
\end{equation}
The light quark $\bar q$ carries the momentum $\eta_1v$ and
$\eta_2 v$ in the amplitude $A$ and $A^\dagger$ respectively. Now
we observe that for $\eta_3\to 0$ the amplitude $A$ and
$A^\dagger$ is finite, while for $\eta_1 \to 0$ and $\eta_2\to 0$
they are divergent respectively.
We expand the amplitude $A$ and $A^\dagger$ respectively in $\eta_1v$ and
$\eta_2 v$ and only keep the leading terms
in the expansion in $\eta_1,\eta_2$ and $\eta_3$, i.e.;
\begin{equation}
A_{ij} = \frac{T_{ij}}{\eta_1} (1+ \cdots), \ \ \
A^\dagger_{ij} = \frac {T^\dagger_{ij}}{\eta_2} (1+\cdots),
\end{equation}
where $T_{ij}$ and $T^\dagger_{ij}$ do not depends on $\eta_1$, $\eta_2$ and
$\eta_3$. Using Eq.(41) the integral over $\eta_3$ and $\omega_3$ can be performed
and we obtain:
\begin{eqnarray}
 d\sigma_R^{(1)} &=&
        \frac{1}{2\hat s}\frac {d^3k}{(2\pi)^3}\int\frac{d^3p_3}{(2\pi)^3}
        (2\pi)^3\delta^3 (p_{T1}+p_{T2}-p_{T3})
        \nonumber \\
   &&   \cdot  \int \frac{d\eta_2}{2\pi}
       \frac{1}{9\eta_1\eta_2} {\rm Tr} T_{ij}(p_1,p_2,m_Qv,p_3)
        {\rm Tr}(\gamma^0 T^\dagger
         (p_1,p_2,m_Qv,p_3)\gamma^0)_{kl} \nonumber\\
      &&\cdot \int d\omega_1 d\omega_2
         e^{+i\eta_2\omega_2-i\eta_1\cdot \omega_1}
        \langle 0\vert \bar q_k(\omega_2 v) h_l(0)  a^\dagger _{H_Q}(k)
       a_{H_Q}(k) \bar h_i(0) q_j(\omega_1v) \vert 0\rangle.
\end{eqnarray}
The Dirac indices of the Fourier transformed matrix element can be decomposed.
In general there are many terms. With the property in Eq.(33) the number of terms
can be reduced greatly and we have only four terms:
\begin{eqnarray}
&& v^0\int \frac{d\omega_1}{2\pi} \frac{d\omega_2}{2\pi}
         e^{-i\eta_1\cdot \omega_1+i\eta_2\cdot \omega_2}
        \cdot \langle 0\vert \bar q_k(\omega_2 v) h_l(0)  a^\dagger _{H_Q}(k)
       a_{H_Q}(k) \bar h_i(0) q_j(\omega_1v) \vert 0\rangle
       \nonumber\\
&& =
   \left ( P_v\right )_{ji}
   \left ( P_v\right )_{lk}
   {\cal W}_1^{(1)}(\eta_1,\eta_2)
   - \left (  \gamma_5 P_v \right )_{ji}
   \left ( P_v\gamma_5\right )_{lk}
    {\cal W}_2^{(1)}(\eta_1,\eta_2)
    \nonumber\\
 &&  - \left (  \gamma_T^\mu P_v\right )_{ji}
   \left ( P_v\gamma_{T\mu}\right )_{lk}
     {\cal W}_3^{(1)}(\eta_1,\eta_2)
   - \left (  \gamma_T^\mu\gamma_5 P_v\right )_{ji}
   \left ( P_v \gamma_{T\mu}\gamma_5\right )_{lk}
        {\cal W}_4^{(1)}(\eta_1,\eta_2)
\end{eqnarray}
where
\begin{equation}
\gamma^\mu_T=\gamma^\mu -v\cdot\gamma v^\mu, \ \ \ \
              P_v =\frac{1+\gamma\cdot v}{2}
\end{equation}
and the functions are defined in the rest frame of $H_Q$. By using
matrix elements in the rest frame the factor $v^0$ in Eq.(43) appears
because of the nonrelativistic normalization of the state. The functions
are:
\begin{eqnarray}
   {\cal W}_1^{(1)}(\eta_1,\eta_2) &=& \frac{1}{4}
   \int \frac{d\omega_1}{2\pi} \frac{d\omega_2}{2\pi}
         e^{-i\eta_1\cdot \omega_1+i\eta_2\cdot \omega_2}
   \cdot \langle 0\vert \bar q(\omega_2 v) h(0)  a^\dagger _{H_Q}
       a_{H_Q} \bar h(0) q(\omega_1v) \vert 0\rangle ,
\nonumber\\
  {\cal W}_2^{(1)}(\eta_1,\eta_2) &=&-\frac{1}{4}
   \int \frac{d\omega_1}{2\pi} \frac{d\omega_2}{2\pi}
         e^{-i\eta_1\cdot \omega_1+i\eta_2\cdot \omega_2}
       \cdot \langle 0\vert \bar q(\omega_2 v)\gamma_5 h(0)  a^\dagger _{H_Q}
       a_{H_Q} \bar h(0) \gamma_5 q(\omega_1v) \vert 0\rangle,
       \nonumber\\
{\cal W}_3^{(1)}(\eta_1,\eta_2) &=& -\frac{1}{12}
   \int \frac{d\omega_1}{2\pi} \frac{d\omega_2}{2\pi}
         e^{-i\eta_1\cdot \omega_1+i\eta_2\cdot \omega_2}
       \cdot \langle 0\vert \bar q(\omega_2 v)\gamma_T^\nu h(0)
        a^\dagger _{H_Q}
       a_{H_Q} \bar h(0) \gamma_{T\nu} q(\omega_1v) \vert 0\rangle,
  \nonumber\\
{\cal W}_4^{(1)}(\eta_1,\eta_2) &=& -\frac{1}{12}
   \int \frac{d\omega_1}{2\pi} \frac{d\omega_2}{2\pi}
         e^{-i\eta_1\cdot \omega_1+i\eta_2\cdot \omega_2}
\cdot \langle 0\vert \bar q(\omega_2v)\gamma_T^\nu\gamma_5 h(0)  a^\dagger _{H_Q}
       a_{H_Q} \bar h(0) \gamma_{T\nu}\gamma_5 q(\omega_1v) \vert 0\rangle,
       \nonumber\\
\end{eqnarray}
In the above definitions the operator $a_{H_Q}$ creates $H_Q$ in its rest frame.
For the color-octet matrix element one has the same decomposition as in Eq.(43)
with the functions ${\cal W}_i^{(8)}(\eta_1,\eta_2,\eta_3)$ with $i=1,\cdots 4$.
The definition of these functions are obtained
by replacing the color matrix $1\otimes 1$ with
$T^a \otimes T^a$ in the above equation.
\par
To express our results as the differential cross-section like $d\sigma/d\hat t$
we need to consider the phase-space integral
\begin{equation}
\int \frac {d^3k}{(2\pi)^3}\cdot\frac{d^3p_3}{(2\pi)^3}
        (2\pi)^3\delta^3 (p_{T1}+p_{T2}-p_{T3}).
\end{equation}
Because $p_1+p_2-k_1-k_3-p_3$ we can insert the identity
\begin{equation}
\int d \eta_1 \delta (v\cdot(p_1+p_2-m_Qv-\eta_1 v-p_3))=1,
\end{equation}
where we neglect $\eta_3$ in Eq.(40) and the momentum $k$ of $H_Q$
is approximated by $k=m_Qv+\eta_1 v$. The $\delta$-function in Eq.(47)
combining those $\delta$-functions in Eq.(46) gives usual $\delta$-functions
$\delta^4(p_1+p_2-k-p_3)$ for momentum conservation, then the phase-space integral
can be calculated as usual. We obtain:
\begin{equation}
\int \frac {d^3k}{(2\pi)^3}\cdot\frac{d^3p_3}{(2\pi)^3}
        (2\pi)^3\delta^3 (p_{T1}+p_{T2}-p_{T3})
  \approx \frac{m_Q}{ 8\pi ^2 \sqrt{\hat s}}
  \int d\hat t d\eta_1 v^0
\end{equation}
where we use $k\approx m_Qv$. Our results contain 8 parameters which represent
nonperturbative effects. These parameters are defined as integrals with
the functions ${\cal W}^{(1,8)}_i(i=1,\cdots 4)$:
\begin{equation}
w^{(1,8)}_i = \frac{1}{m_Q^3}\int \frac{d\eta_1}{\eta_1}\frac{d\eta_2}{\eta_2}
                 {\cal W}_i^{(1,8)}(\eta_1,\eta_2).
\end{equation}
These parameters are dimensionless. At first look, they may scale as
$\Lambda_{QCD}^3/m_Q^3$. But, the dominant contribution of the integrals
comes from the region where $\eta_1 \sim \eta_2 \sim \Lambda_{QCD}$,
reflecting the fact that the light antiquark $\bar q$ carries
a small fraction of the momentum of $H_Q$. This results in that
the parameters $w^{(1,8)}_i(i=1,2,3,4)$ scale as $\Lambda_{QCD}/m_Q$.
With these parameters our results
for the process $g+\bar{q}\rightarrow H_Q+X$ are:
\begin{equation}
 \frac {d\sigma_R^{(i)}}{d\hat t}(\hat s,\hat t)
  =\frac {\pi\alpha_s^3 m_Q^2}{ \hat{s}^2}
\sum\limits_{j=1}^4 w_j^{(i)} {\cal B}_j^{(i)},
\end{equation}
where
\begin{eqnarray}
{\cal B}_1^{(1)}&=&{\cal B}_2^{(1)}=\frac{4}{81\hat{s}}\Big\{
\frac{m_Q^2\Big[127 T^2-16\hat{s} T-64\hat{s}^2\Big]}{T^3}
-\frac{64 U}{\hat{s}}-\frac{16 m_Q^4 \hat{s}}{U T^2}(1-\frac{8U}{T})\Big\},\nonumber\\
{\cal B}_3^{(1)}&=&{\cal B}_4^{(1)}=\frac{1}{81\hat{s}}\Big\{-
\frac{m_Q^2}{T}\Big[ 19 + \frac{4\,T^2}{U^2}-\frac{28 T}{U} +
  \frac{368 U}{T} - \frac{64 U^2}{T^2} \Big] \nonumber\\
&&-\frac{64 U}{\hat{s}}\Big[1+\frac{2U^2}{T^2}\Big]
-\frac{48 m_Q^4\hat{s} (8\hat{s}+9 T)}{T^3 U}
\Big\},
\end{eqnarray}

\begin{eqnarray}
{\cal B}_1^{(8)}&=&{\cal B}_2^{(8)}=\frac{2}{27\hat{s}}\Big\{
\frac{m_Q^4\hat{s} (8\hat{s}^2+8\hat{s} T+9 T^2)}{T^3\,U^2}
-\frac{m_Q^2}{2T}\Big[ 79-\frac{18 T^2}{U^2} + \frac{14 U}{T} +
\frac{8\,U^2}{T^2}\Big] \nonumber\\
&&-\frac{4\hat{s}^2 - \hat{s} T + 4 T^2}{\hat{s}\,U} \Big\}, \nonumber\\
{\cal B}_3^{(8)}&=&{\cal B}_4^{(8)}= \frac{2}{27\hat{s}}\Big\{
-\frac{U}{\hat{s}}\Big[
22+\frac{9T}{U}+\frac{18U}{T}+\frac{9T^2}{U^2}+\frac{8U^2}{T^2}\Big]\nonumber\\
&&-\frac{m_Q^2}{2T}\Big[233+\frac{266 T^2}{U^2}+\frac{316 T}{U}+
\frac{10 U}{T}-\frac{8 U^2}{T^2}\Big]+
\frac{3m_Q^4\hat{s}}{U T^2}\Big[8+\frac{9 T}{U}+\frac{8 U}{T}\Big]\Big\},
\end{eqnarray}
with
\begin{equation}
T=\hat{t}-m_Q^2,~~~~~~~U=-\hat{s}-T.
\end{equation}
\par
Similarly for the process $\gamma+\bar{q}\rightarrow H_Q+X$, we have
\begin{equation}
 \frac{d\tilde{\sigma}_R^{(i)}}{d\hat t}(\hat s,\hat t) =
\frac{2\pi \alpha \alpha_s^2 e_Q^2 m_Q^2}{3 \hat{s}^2}
{\cal C}^{(i)}
 \sum\limits_{j=1}^5
w_j^{(i)} \tilde{{\cal B}}_j,
\end{equation}
where
\begin{eqnarray}
\tilde{\cal B}_1&=&\tilde{\cal B}_2=\frac{1}{4\hat{s}}\Big\{
\frac{2m_Q^4\hat{s}^3}{T^3 U^2}(1+\kappa\frac{T}{\hat{s}})
+\frac{m_Q^2\hat{s}}{U^2}\Big[
\frac{2(1+\kappa)\hat{s}}{T}+4\kappa+
\frac{\kappa^2 T}{\hat{s}}-\frac{\hat{s}^3}{T^3} \Big]
-\frac{(\hat{s}+\kappa \hat{s} T)^2}{\hat{s}U}
\Big\}, \nonumber\\
\tilde{\cal B}_3&=&\tilde{\cal B}_4=\frac{1}{4\hat{s}}\Big\{
\frac{m_Q^2 \hat{s}}{U^2}\Big[4\kappa (3+\kappa)+\frac{s^3}{T^3}
+\frac{4 (2+\kappa) s^2}{T^2}+\frac{2 (3+7 \kappa) s}{T}+\frac{3 \kappa ^2 T}{s}\Big]
 \nonumber\\
&&+\frac{6m_Q^4 \hat{s}^2 (\hat{s} + \kappa T) }{T^3 U^2}
-\frac{\hat{s}}{U}\Big[ (1+\frac{\kappa T}{\hat{s}})^2 (1+\frac{2 U^2}{T^2})\Big]\Big\},
\end{eqnarray}
with ${\cal C}^{(1,8)}$ given in Eq.(25). In above results polarizations of initial
states are averaged.
\par
Our results contain 8 parameters in general which are defined as integrals
with matrix elements of HQET. They are universal, i.e.,
they do not depend on a specific process.
With our results one can check that the ratio is a constant:
\begin{equation}
  \frac{d\sigma (g+\bar q\to H_Q+X)}{d\sigma(g+g \to H_Q +X)}\vert_{
   \hat t\to -\vert \hat t\vert_{min}}
   \sim {\rm const.},
\end{equation}
i.e., the ratio does not depend on $\hat s$, hence the contribution from
quark recombination is not suppressed by inverse of certain power of $\hat s$,
and it can give a significant contribution. This contribution is only
suppressed by the parameters given in Eq.(49).
\par
The 8 parameters can be effectively reduced to 4 because of the property of
${\cal B}_1^{(1,8)}={\cal B}_2^{(1,8)}$ and ${\cal B}_3^{(1,8)}={\cal B}_4^{(1,8)}$.
One can show this property without explicit calculation.
With the expansion in Eq.(41) the last diagram in Fig.1. will not contribute.
Taking any contribution from the first 4 diagrams in Fig.1, e.g., the interference
between amplitudes from Fig.1a and Fig.1b,
the coefficient of $w_1^{(1)}$ can be written as a trace:
\begin{equation}
{\cal B}_1^{(1)} \sim {\rm Tr}\left \{
  \cdots I \gamma\cdot p_2 I \cdots \right \},
\end{equation}
where $\cdots$ denote polynomials of products of $\gamma$ matrices, $I$ denotes
$4\times 4$ unit matrix coming from Eq.(43). Since we do not observe the polarization
of initial hadrons, the spin of the initial light quark $\bar q(p_2)$
is averaged, it results in the factor $\gamma\cdot p_2$.
The same leads to the coefficient
of $w_2^{(1)}$:
\begin{equation}
{\cal B}_2^{(1)} \sim {\rm Tr}\left \{
  \cdots (-\gamma_5) \gamma\cdot p_2 \gamma_5  \cdots \right \},
\end{equation}
it is clear that both coefficient are same. If initial hadrons are polarized,
the coefficients are in general not the same. To illustrate this, we consider
the case where only the hadron containing the light quark $\bar q(p_2)$ is
transversally polarized with the polarization vector $S_T^\mu$ and
another initial hadron is unpolarized.
In this case spin-dependent parts of perturbative coefficients, denoted as
${\cal B}_{si}^{(1,8)}(i=1,2,3,4)$,
can be calculated by replacing $\gamma\cdot p_2$ with
$i\sigma_{\mu\nu}\gamma_5 p_2^\mu S_T^\nu$,
the contribution to the spin-dependent part of the cross section
is a convolution with the so-called transversity distribution function
instead of usual parton distribution functions\cite{JaJi}.
With the replacement one can easily find that spin-dependent parts of
perturbative coefficients have the properties
${\cal B}_{s1}^{(1,8)}=-{\cal B}_{s2}^{(1,8)}$ and
${\cal B}_{s3}^{(1,8)}=-{\cal B}_{s4}^{(1,8)}$,
in contrast to the spin-independent part. Hence, if the spin-dependent
parts of perturbative coefficients are not zero, we will have 8 nonperturbative
parameters which lead to different contributions to the cross section.
With symmetries one can also identify when these spin-dependent parts
are zero in this case. With the rotational symmetry, the spin-dependent
part of differential cross-sections consists of two parts, one is
proportional to ${\bf S_T} \cdot{\bf k}$, while another is proportional
to ${\bf S_T}\cdot ({\bf P }\times {\bf k})$, where ${\bf P}$ is the
three momentum of the polarized hadron. If the parity is conserved,
the term proportional to ${\bf S_T} \cdot{\bf k}$ is zero. With the
time-reversal symmetry one can show that the term proportional to
${\bf S_T}\cdot ({\bf P }\times {\bf k})$ is zero if
no absorptive parts in amplitudes. Further, one can show
with arguments from chirality that the two terms are proportional
to $m_Q$. Therefore, in the case considered
here, we will have effectively 4 parameters only provided that there is no
parity-violating interaction and absorptive part in partonic processes.
Since parity-violating interactions can be involved and an
absorptive part can also appear beyond tree-level, we have in general
8 parameters in differential cross-section. Similarly, one can also show
that the 8 parameters lead to different contributions in the case
if all initial hadrons are polarized. It should be noted that
in the above discussion the polarization of the final hadron, if
it is not spin-less, is summed.
\par
Because all these
parameters are unknown, a numerical prediction can not be made.
As an estimation, one may use vacuum saturation
for $w_i^{(1)}(i=1,\cdots,4)$, i.e., the color singlet parts, to relate them
to wave functions of $H_Q$. However, the approximation of
the vacuum saturation is not well established because $\bar q$ is
a light quark, and also the approximation definitely does not apply
for the color octet part, i.e., for  $w_i^{(8)}(i=1,\cdots,4)$.
With the definitions
of these parameters one may use nonperturbative methods like QCD sum rules
to study them, or they can be extracted from experiment. Once their numerical
values are known, numerical predictions can be made with our results. In this work
we do not make an attempt to fit experimental results with our results
for determining these parameters,  because in \cite{B1,B2,B3} it is shown
that one can already describe the leading particle effect
observed in experiment by keeping contributions with $w_{2,3}^{(1,8)}$ and neglecting
other contributions. The relation between the parameters in \cite{B2} and ours can be
identified:
\begin{eqnarray}
\rho_1(Q\bar q(^1S_0) \to H_Q) &=&\frac{1}{2} w_2^{(1)}, \ \ \ \ \
\rho_1(Q\bar q(^3S_1) \to H_Q) =\frac{1}{2} w_3^{(1)},
\nonumber\\
\rho_8(Q\bar q(^1S_0) \to H_Q) &=& \frac{3}{8} w_2^{(8)}, \ \ \ \ \
\rho_8(Q\bar q(^3S_1) \to H_Q) =\frac{3}{8} w_3^{(8)}.
\end{eqnarray}
\par
It should be noted that in the contributions related to $w_{2,3}^{(1,8)}$
the pair $Q \bar q$ is in $^1S_0$ and $^3S_1$ state, respectively. Our results
of these contributions to the cross-section are exactly the same as those given in \cite{B2,B3}.
From our general
analysis one can see that the pair $Q\bar q$ can form  a scalar, a pseudoscalar,
a vector, and a pseudovector state. All of them can lead to contributions
to the leading particle effect, which can be predicted with the above results
by using Eq.(28) or Eq.(29). If the light quark $\bar q$ is replaced with a heavy quark,
one can use NRQCD factorization\cite{NRQCD} to make prediction, then
at leading order of the factorization, the $Q\bar q$ can only form a pseudoscalar
or vector state, i.e., $^1S_0$- and $^3S_1$ state. In this approach motivated
by NRQCD factorization only
terms with $w_{2,3}^{(1,8)}$ will remain, while other terms are at higher
order in the approach. This approach is used in \cite{B1,B2,B3}. With this approach
for polarized beams one will still have 4 parameters in contrast to our results containing
8 parameters.
\par

\par\vskip20pt\noindent
{\bf 4. Summary}
\par\vskip20pt
In this work we have made a general analysis for
the effect of quark recombination in inclusive production of a heavy meson.
Through these effects the leading particle effect observed in experiment
can be explained. In the contributions of quark recombination
to differential cross section we factorize perturbative- and nonperturbative
parts in two different cases. One is
for large transverse momentum of the produced hadron,
where the transverse momentum
is so large that any quark mass can be neglected. In this case we find the
nonperturbative effect
of quark recombination is parameterized by six functions defined with four quark
operators at twist-4. Four of these functions contribute
to leading particle effect, if the initial state is unpolarized.
The observed effect of leading particle is of charmed hadron, one can
take the charm quark as a heavy quark and perform the factorization
with HQET. In this case the nonperturbative part consists of 8 parameters,
which are defined as integrals of matrix elements with 4-quark operators
in HQET. These parameters represent the transition of a quark pair $Q\bar q$
in different states into the produced hadron,
the pair can be in color singlet or color octet
state and forms different states as scalar, pseudoscalar, vector and pseudovector
state.
The results obtained here hold not only
for large transverse momentum, but also for small transverse momentum.
Our results are different than those obtained with a factorization
motivated by NRQCD\cite{B1,B2,B3},
in which the pair only forms pseudoscalar or
vector state, although for unpolarized beams both
results contain the same number of nonperturbative parameters.
The perturbative part in both types of factorization
are calculated at tree-level for hadro- and photoproduction.
\par
The nonperturbative functions or parameters are unknown yet. This fact
prevents us from a numerical prediction for leading particle effect. But
they can be studied by nonperturbative methods like QCD sum rules or models,
or they can be extracted from experimental results.
In this work we have not tried to extract the eight parameters from experimental
results, because it is shown in \cite{B1,B2,B3}
that one can already describe the leading particle effect
observed in experiment only by keeping contributions of $w_{2,3}^{(1,8)}$,
and contributions of $w_{1,4}^{(1,8)}$ in our result is proportional
to the contributions  of $w_{2,3}^{(1,8)}$, respectively.
Hence,
with effects of all possible states of $Q\bar q$ the leading particle effect
can also be generated and can be described. It should be noted
that the leading particle effect of charmed baryon can be also
analyzed in a similar way presented in this work, results will
be published elsewhere\cite{Next}.

\par\vskip20pt\noindent
{\bf Acknowledgements}
\par
We would like to thank Prof. E. Braaten, Prof. Y.Q. Chen, Dr. Yu Jia and
Prof. J.S. Xu  for discussions,   Z.G.Si would like
to thank members in the
theoretical physics group of Shandong University for their useful discussions.
The work of J.P. Ma is supported by National Nature
Science Foundation of P. R. China with the grand No. 19925520.
\vfil\eject

\end{document}